\documentclass{llncs}

\usepackage[latin1]{inputenc}
\usepackage{bbm}
\usepackage[T1]{fontenc}

\newcommand{\rt}{\Rightarrow}
\newcommand{\Lra}{\Leftrightarrow}
\newcommand{\sm}{\mbox{ }}
\newcommand{\vm}[1]{\vspace{#1em}}
\newcommand{\hx}[1]{\hspace{#1ex}}
\newcommand{\nq}{\neq}
\newcommand{\eq}{\equiv}
\newcommand{\fa}{\forall}
\newcommand{\ex}{\exists}
\newcommand{\mx}{\mbox}
\newcommand{\nt}{\neg}
\newcommand{\we}{\wedge}
\newcommand{\ca}{\cap}
\newcommand{\tm}{\times}
\newcommand{\fc}{\frac}
\newcommand{\pe}{\prec}
\newcommand{\su}{\subset}
\newcommand{\ld}{\ldots}
\newcommand{\ns}{\mathbbmss{N}}
\newcommand{\rl}{\mathbbmss{Q}}
\newcommand{\rs}{\mathbbmss{R}}
\newcommand{\zs}{\mathbbmss{Z}}
\newcommand{\lj}{{\sf CEDRIC}}
\newcommand{\uc}{{\sf CNAM}}
\newcommand{\er}{{\sf CPR}}
\newcommand{\pn}{{\sf LCR}}
\newcommand{\iv}{{\sf LIPN}}
\newcommand{\cq}{{\sf Coq}}
\newcommand{\hl}{{\sf HOL}}
\newcommand{\hq}{{\sf HOL90}}
\newcommand{\ie}{{\sf Isabelle}}
\newcommand{\mr}{{\sf Mizar}}

\newcommand{\dc}{{\sf David.Delahaye@cnam.fr}}
\newcommand{\dr}{{\sf http://cedric.cnam.fr/\~{}delahaye/}}
\newcommand{\mc}{mayero@lipn.univ-paris13.fr}
\newcommand{\mh}{http://www-lipn.univ-paris13.fr/\~{}mayero/}

\begin{document}

\title{Diophantus' 20th Problem and\\Fermat's Last Theorem for $n=4$\\\vm{1}
{\normalsize Formalization of Fermat's Proofs\\in the \cq{} Proof Assistant}}

\author{David~Delahaye\inst{1} \and Micaela~Mayero\inst{2}}

\institute{\er{} (\lj{}), \uc{}, Paris, France\\
\email{\dc{}}\\
\texttt{\dr{}}\vm{1}
\and \pn{} (\iv{}), Université Paris Nord (Paris~13),\\
Villetaneuse, France\\
\email{\mc{}}\\
\texttt{\mh{}}}

\maketitle

\begin{abstract}
We present the proof of Diophantus' 20th problem (book~VI of Diophantus'
{\em Arithmetica}), which consists in wondering if there exist right triangles
whose sides may be measured as integers and whose surface may be a square.
This problem was negatively solved by Fermat in the 17th century, who used the
{\em wonderful} method ({\em ipse dixit} Fermat) of infinite descent. This
method, which is, historically, the first use of induction, consists in
producing smaller and smaller non-negative integer solutions assuming that one
exists; this naturally leads to a {\em reductio ad absurdum} reasoning because
we are bounded by zero. We describe the formalization of this proof which has
been carried out in the \cq{} proof assistant. Moreover, as a direct and no
less historical application, we also provide the proof (by Fermat) of Fermat's
last theorem for $n=4$, as well as the corresponding formalization made in
\cq{}.
\end{abstract}

\section{Introduction}

Diophantus of Alexandria (c. {\sc ad} 250) was a Greek mathematician whose life
is little known but who wrote the 13~books of a collection called {\em
Arithmetica}~\cite{Dio-Ari}. Diophantus is usually considered to be the father
of Algebra, and his books consider more than 130~problems (most of which have
been solved) of first and second order leading to equations whose roots are
either integer or fractional. Until~1972, only 6~books of this collection had
been retrieved (in the 15th century in Italy by Regiomontanus) when 4~other
books were found in Iran. The collection was translated in the 16th century by
Wilhelm~Holtzmann (also known as Xylander) at Heidelberg (in Germany) and
completed (in France) in Latin by Claude-Gaspard~Bachet~De~Méziriac.
Diophantus' work had a significant influence on Arabic mathematicians but also
on western (and essentially French) mathematicians like Viete and Fermat. In
the 17th century, reading Bachet's translation (now lost) of book~VI (related
to propositions over right triangles whose sides are measured as integers),
Pierre~Simon~de~Fermat (1601-1665)~\cite{PSF22} was interested, amongst others,
in the following problem (20th problem): can a right triangle whose sides are
measured as integers have a surface measured as a square? Formally, this is
equivalent to knowing if there exist four non-zero integers $x$, $y$, $z$ and
$t$ s.t.:
$$x^2+y^2=z^2\mx{ and }xy=2t^2\mx{.}$$

We know that the first equation has an infinity of solutions (for example, $3$,
$4$ and $5$, etc), called Pythagorean triples~\cite{Euc-Ele} (for they measure
the sides of a right triangle and verify Pythagoras' relation), but with the
condition over the surface the problem is a little more difficult so that
Fermat answered this question negatively~\cite{Dio-proof} using a {\em
wonderful} method (the word was applied by Fermat himself): the infinite
descent~\cite{PSF22,AW84,JI63}. This method is based on the fact that there
does not exist any strictly decreasing non-negative integer sequence. Thus,
starting from a lemma characterizing Pythagorean triples, Fermat's idea
consists in re-expressing the problem with (strictly) smaller non-negative
integers. More precisely, Fermat concludes his proof as follows (quotation of
the original text~\cite{PSF22} in modern French):

{\small \em
\begin{quote}
Si on donne deux carrés dont la somme et la différence sont des carrés, on
donne par là même, en nombres entiers, deux carrés jouissant de la même
propriété et dont la somme est inférieure.\\

Par le même raisonnement, on aura ensuite une autre somme plus petite que celle
déduite de la première, et en continuant indéfiniment, on trouvera toujours des
nombres entiers de plus en plus petits satisfaisant aux mêmes conditions. Mais
cela est impossible, puisqu'un nombre entier étant donné, il ne peut y avoir
une infinité de nombres entiers qui soient plus petits.
\end{quote}}

which means that given two squares $m^2$, $n^2$ s.t. $m^2+n^2$ and $m^2-n^2$
are also squares, we can find two squares $m'^2$, $n'^2$ with the same
properties s.t.\linebreak{}$m'^2+n'^2<m^2+n^2$. Re-applying the process
infinitely, we always find smaller non-negative integers (w.r.t. $m^2+n^2$),
which is impossible because we are\linebreak{}bounded by zero.

This proof is worth being formalized in a theorem prover for several reasons.
First, this is a {\em nice} mathematical proof in the sense that it is rather
short (without, nonetheless, being trivial) and uses an original method
(infinite descent). Actually, it can be shown that the descent is equivalent to
Noetherian induction and even if it is difficult to consider induction
reasoning as original these days, it is more the expression of this induction
(making it possible to establish universally false propositions) which is
interesting here (this method has not been greatly formalized or even used in
deduction systems). This provides an additional interest to Fermat's proof and
to this work since this is the first use of induction in the history of
Mathematics. Moreover, beyond the fact that adding this new theorem contributes
a little more to the formalization of Mathematics on a computer, the true
challenge is certainly the development of the application of the method itself
(which can vary widely from one problem to another\footnote{For example, using
this method to prove Fermat's last theorem for $n=4$ may be considered as
rather elementary, whereas the proof of Leonhardt~Euler for $n=3$ ruins any
hope, for Christian~Goldbach (his friend and boss), of using such a method to
find a general proof for this theorem.}). Finally, this proof has a high re-use
potential. Fermat's last theorem~\cite{SS02,AW84,DS93,JI63} (there do not exist
non-zero integers $x$, $y$ and $z$ s.t. $x^n+y^n=z^n$ for $n>2$) can be easily
deduced for $n=4$ (also proved by Fermat) from the proof of Diophantus' 20th
problem and we also provide the proof in this paper as well as its
formalization. Infinite descent is also used to prove Fermat's last theorem for
$n=3$ (probably first proved by Fermat and later by Leonhardt~Euler and
Karl~Friedrich~Gauss independently), $n=5$ (proved by Adrien-Marie~Legendre and
Lejeune~Dirichlet using Sophie~Germain's work), $n=7$ (proved by Gabriel~Lamé)
and $n=14$ (proved by Dirichlet). More generally, as claimed in~\cite{AW84},
the infinite descent method is the method {\em par excellence} in number theory
and in Diophantine analysis in particular.

As a theorem prover, we chose to use the \cq{} proof
assistant~\cite{CoqDocWeb8.0} (V8.0). Despite the fact that \cq{} is usually
not considered to be one of the most mathematician-friendly theorem provers
(essentially due to its proof style, i.e. the proofs are expressed in a
procedural way which may seem unnatural for mathematician users, and probably a
not high enough level of automation, i.e. the system may be, in some cases, not
strong enough to deduce automatically theorems from others whereas it seems
rather easy to do so by hand), our choice was motivated both by recent
improvements regarding concrete syntax, in particular for arithmetic, and by a
fairly sufficient degree of automation for the problem we wanted to formalize
(actually, only ring simplifications were needed in our development).

In this paper, we present an {\em informal} (but rigorous) sketch of Fermat's
proofs for Diophantus' 20th problem and Fermat's last theorem for $n=4$, as it
would be described in a usual Mathematics book. Next, we give details regarding
the formalization of this proof emphasizing the difficult points (essentially
the lemmas related to Pythagorean triples and the descent) and the solutions we
provided.

\section{Mathematical proof sketch}
\label{mth_prf}

As said in the introduction, we want to prove that there do not exist right
triangles whose sides are measured as integers and the surface as a square.
This means that there do not exist four non-zero natural numbers (the theorem
is also true for integers) $x$, $y$, $z$ and $t$ s.t.:
$$x^2+y^2=z^2\mx{ and }xy=2t^2\mx{.}$$

The proof starts looking for a characterization of Pythagorean triples, i.e.
the set of triples of natural numbers $x$, $y$ and $z$ verifying $x^2+y^2=z^2$.

In the following, $\ns{}$ denotes the set of natural numbers (considering that
$0\in{}\ns{}$), i.e. the set of non-negative integers, and $\ns{}^*$ is the set
of natural numbers except $0$, i.e. the set of positive integers.

\subsection{Pythagorean triples}
\label{pyt_tri}

Historically, Pythagorean triples (also called Pythagorean triads) were studied
by Euclid of Alexandria in his {\em Stoicheion}~\cite{Euc-Ele} ({\em The
Elements}). But, as can be seen in~\cite{AW84}, a Babylonian tablet
(Plimpton~322; c. BC 1900-1600) already contained the computation of fifteen
Pythagorean triples, which tends to prove that such triples were at least known
long before Euclid and may even have been calculated according to some rules.
The set of Pythagorean triples can be characterized by theorem~\ref{ptr_thm}
below. The proof, we provide, uses a geometrical point of view and consists in
locating the rational points of the unit circle. This proof is described
in~\cite{KD94} and is far different from the usual proofs that can be found
in~\cite{HW79} or~\cite{DS93}.

\begin{theorem}[Pythagorean triples]
\label{ptr_thm}
Let ${\cal S}$ be the set of Pythagorean triples and defined as
${\cal S}=\{\sm{}(a,b,c)\sm{}|\sm{}a,b,c\in{}\ns{}\sm{}and\sm{}a^2+b^2=c^2
\sm{}\}$. Let ${\cal T}$ be the set defined as follows:
$$\begin{array}{lll}
{\cal T}=\{ & (m(q^2-p^2),2mpq,m(p^2+q^2)),\\
& (2mpq,m(q^2-p^2),m(p^2+q^2))| & m,p\in{}\ns{},q\in{}\ns{}^*,p\le{}q,\\
&& p\mx{ and }q\mx{ relatively prime},\\
&& p\mx{ and }q\mx{ have distinct parities}\}\mx{.}\end{array}$$

Then ${\cal S}={\cal T}$.
\end{theorem}

\begin{proof}
We denote $C=\{(x,y)\in{}\rs{}^2|x^2+y^2=1\}$, the unit circle and, for
$r\in{}\rs{}$, $D_r=\{(x,y)\in{}\rs{}^2|y=r(x+1)\}$. The proof is made in
6~steps:\\

\noindent{}{\bf Step~1:} given a Pythagorean triple $(a,b,c)$, which is not
$(0,0,0)$, there exists a corresponding point $(\fc{a}{c},\fc{b}{c})$ of the
unit circle. As $c>0$, we can divide by $c^2$:
$(\fc{a}{c})^2+(\fc{b}{c})^2=1$, which verifies the unit circle equation.
Conversely, given a point $(\fc{a}{c},\fc{b}{c})$ of the unit circle, there
exists an infinity of corresponding Pythagorean triples $(ma,mb,mc)$, for
$m\in{}\ns{}$. We have $(\fc{a}{c})^2+(\fc{b}{c})^2=1$ and we can multiply by
$m^2c^2$ obtaining: $(ma)^2+(mb)^2=(mc)^2$.\\

\noindent{}{\bf Step~2:} the set $C\ca{}D_r$ has two points. To find these
points, we have to solve the following system:
\begin{equation}
\left\{\begin{array}{l}
y^2=1-x^2\\
y=r(x+1)
\end{array}\right.
\label{int_sys}
\end{equation}

Thus, $x$ must be solution of the following equation:
$$(1+r^2)x^2+2r^2x+r^2-1=0$$

The solutions are $-1$ and $\fc{1-r^2}{1+r^2}$. Using the second equation
of~(\ref{int_sys}), we obtain the two solutions
$\{(-1,0);(\fc{1-r^2}{1+r^2},\fc{2r}{1+r^2})\}$. We notice that the second
point is non-negative for $0\le{}r\le{}1$.\\

\noindent{}{\bf Step~3:} now, given $M\in{}C$, we can show that the coordinates
of $M$ are rational iff there exists a rational $r$ s.t. $M\in{}C\ca{}D_r$.
First, let us suppose that we have $r\in{}\rl{}$ with $M\in{}C\ca{}D_r$.
We have two possibilities: either $M=(-1,0)$, which is trivially rational, or
$M=(\fc{1-r^2}{1+r^2},\fc{2r}{1+r^2})$, where the coordinates are rational
fractions (quotients of polynomials) in $r\in{}\rl{}$, thus also in $\rl{}$.\\

Conversely, let us suppose the coordinates $(x,y)$ of $M$ are rational. We have
two cases: either $M=(-1,0)$ and $M$ is in $M\in{}C\ca{}D_r$, for all
$r\in{}\rl{}$, or else $M\nq{}(-1,0)$ and we take $r=\fc{y}{x+1}$ (which is a
rational), $M$ is in $C$ by hypothesis as well as in $D_r$ by construction of
$r$.\\

\noindent{}{\bf Step~4:} the points of $C$ with non-negative rational
coordinates are given by the set $\{(\fc{1-r^2}{1+r^2},\fc{2r}{1+r^2})\}$, with
$r\in{}\rl{}\ca{}[0;1]$ (steps~2 and~3). Taking $r=\fc{p}{q}$, with
$p\in{}\ns{}$, $q\in{}\ns{}^*$, $p\le{}q$ and $p$, $q$ relatively prime
(irreducible fraction), the set of points of $C$ with non-negative rational
coordinates is the following:
$${\cal W}=\{(\fc{q^2-p^2}{p^2+q^2},\fc{2pq}{p^2+q^2})|p\in{}\ns{},
q\in{}\ns{}^*,p\le{}q,p\mx{ and }q\mx{ relatively prime}\}$$

\noindent{}{\bf Step~5:} It is not possible to derive a characterization of
Pythagorean triples from ${\cal W}$ because the rational points of the unit
circle must be expressed with irreducible fractions. Hence, let us consider the
set ${\cal W}'$ defined as follows:
$$\begin{array}{lrl}
{\cal W}'=\{ & (\fc{q^2-p^2}{p^2+q^2},\fc{2pq}{p^2+q^2}),
(\fc{2pq}{p^2+q^2},\fc{q^2-p^2}{p^2+q^2})~|\\
& p\in{}\ns{},q\in{}\ns{}^*,p\le{}q, & p\mx{ and }q\mx{ relatively prime},\\
&& p\mx{ and }q\mx{ have distinct parities}\}
\end{array}$$

Let us show that ${\cal W}={\cal W}'$. First, let us consider the inclusion
${\cal W}\su{}{\cal W}'$: given a point
$x=(\fc{q^2-p^2}{p^2+q^2},\fc{2pq}{p^2+q^2})\in{}{\cal W}$, since $p$ and $q$
are relatively prime, either $p$ and $q$ have distinct parities, or they are
both odd. In the former case, we have trivially $x\in{}{\cal W}'$. In the
latter case, let us look for $p'$ and $q'$ s.t.:
\begin{equation}
\fc{q^2-p^2}{p^2+q^2}=\fc{2p'q'}{p'^2+q'^2}\hx{4}\mx{and}\hx{4}
\fc{2pq}{p^2+q^2}=\fc{q'^2-p'^2}{p'^2+q'^2}
\label{sym_sys}
\end{equation}

which leads to the solutions $p'=\fc{q-p}{2}$ and $q'=\fc{p+q}{2}$. These
solutions are both integers since $p$ and $q$ are both odd. We have $p'+q'=q$
and $q'-p'=p$; since $p$ and $q$ are relatively prime, $p'$ and $q'$ are
relatively prime (knowing that if $m+n$ and $m-n$ are relatively prime then $m$
and $n$ are relatively prime). Since $p$ and $q$ are both odd, we have
$p=2k+1$, $q=2k'+1$ and we obtain $p'=k'-k$, $q'=k+k'+1$. Considering all the
cases w.r.t. the parities of $k$ and $k'$, we easily verify that $p'$ and $q'$
have distinct parities. Thus, $x\in{}{\cal W}'$.\\

Conversely, let us prove the inclusion ${\cal W}'\su{}{\cal W}$. Given a point
$x\in{}{\cal W}'$, either $x=(\fc{q^2-p^2}{p^2+q^2},\fc{2pq}{p^2+q^2})$ or
$x=(\fc{2pq}{p^2+q^2},\fc{q^2-p^2}{p^2+q^2})$. In the former case, $x$ is
trivially in ${\cal W}$. In the latter case, we have to solve the
system~(\ref{sym_sys}), which leads to the solutions $p'=q-p$ and $q'=p+q$.
These solutions have distinct parities (using the conditions over $p$ and $q$
together with proposition~\ref{prp_no1} in subsection~\ref{inf_des}). Thus,
$x\in{}{\cal W}$ and we have shown that ${\cal W}={\cal W}'$.\\

\noindent{}{\bf Step~6:} We have to show that ${\cal S}={\cal T}$. Given
$(a,b,c)\in{}{\cal S}$, $(\fc{a}{c},\fc{b}{c})$ is a point of $C$ (step~1),
which can be written as $(\fc{q^2-p^2}{p^2+q^2},\fc{2pq}{p^2+q^2})$ or
$(\fc{2pq}{p^2+q^2},\fc{q^2-p^2}{p^2+q^2})$ (step~5). The two fractions
$\fc{q^2-p^2}{p^2+q^2}$ and $\fc{2pq}{p^2+q^2}$ are irreducible (because $p$
and $q$ are relatively prime and have distinct parities), so $c$ is a multiple
of $p^2+q^2$. Setting $c=m(p^2+q^2)$, we obtain the triple
$(a,b,c)=(m(q^2-p^2),2mpq,m(p^2+q^2))$ or
$(a,b,c)=(2mpq,m(q^2-p^2),m(p^2+q^2))$. Thus, ${\cal S}\su{}{\cal T}$.\\

Given a triple $(a,b,c)\in{}{\cal T}$, either
$(a,b,c)=(m(q^2-p^2),2mpq,m(p^2+q^2))$ or
$(a,b,c)=(2mpq,m(q^2-p^2),m(p^2+q^2))$. In both cases, we only have to verify
that we have a Pythagorean triple (by computation), i.e.:
$$\begin{array}{cl}
(m(q^2-p^2))^2+(2mpq)^2 & =(2mpq)^2+(m(q^2-p^2))^2\\
& =m^2(q^4+p^4-2p^2q^2+4p^2q^2)\\
& =m^2(p^2+q^2)^2=(m(p^2+q^2))^2
\end{array}$$

Thus, ${\cal T}\su{}{\cal S}$ and we have shown that ${\cal T}={\cal S}$.
\end{proof}

\subsection{Infinite descent}
\label{inf_des}

For this proof, which is an application of the infinite descent
method~\cite{PSF22,AW84,JI63}, we essentially used~\cite{Dio-proof}, but it is
also described in~\cite{JI63}. This proof can also be found in~\cite{HW79}
and~\cite{DS93}, integrated into the proof of Fermat's last theorem for $n=4$.

Using theorem~\ref{ptr_thm}, we can express the surface of the right triangle
as:
\begin{equation}
\fc{xy}{2}=k^2pq(q^2-p^2)
\label{dio_nwf}
\end{equation}

with $k,p\in{}\ns{}$, $q\in{}\ns{}^*$, $p\le{}q$, $p$, $q$ are relatively prime
and have distinct parities.

Thus, Diophantus' 20th problem is equivalent to asking:
$$\mx{Can }pq(q^2-p^2)\mx{ be a square?}$$

\subsubsection{Preliminaries}

Here are some preliminary propositions (related to properties regarding
relatively prime integers and squares) we will have to use when building the
infinite descent proof (to save space, we do not provide the proofs of these
rather basic notions):

\begin{proposition}
\label{prp_no1}
Given $m,n\in{}\ns{}$ s.t. $n<m$, if $m$, $n$ are relatively prime and have
distinct parities then $m+n$ and $m-n$ are relatively prime.
\end{proposition}

\begin{proposition}
\label{prp_no2}
Given $m,n\in{}\ns{}$ s.t. $n\le{}m$, if $m$, $n$ are relatively prime then
$m^2$, $n^2$ are relatively prime and $m$, $n$, $m^2-n^2$ are relatively prime.
\end{proposition}

\begin{proposition}
\label{prp_no3}
Given $m,n\in{}\ns{}$, if $m^2$, $n^2$ are relatively prime then $m$, $n$ are
relatively prime.
\end{proposition}

\begin{proposition}
\label{prp_no4}
Given the sequence $(u_n)$ over $\ns{}$, if $u_0,u_1,\ld{},u_n$ are relatively
prime and $u_0\tm{}u_1\tm{}\ld{}\tm{}u_n$ is a square then $u_0,u_1,\ld{},u_n$
are squares.
\end{proposition}

We also recall Gauss's theorem (we do not give the proof again because this
is quite an usual theorem, which, in particular, is already part of the \cq{}
standard library):

\begin{theorem}[Gauss's theorem]
\label{gss_thm}
Given $a,b\in{}\ns{}$, if $d$ divides $ab$ and if $a$, $d$ are relatively prime
then $d$ divides $b$.
\end{theorem}

To make the dependencies between the previous propositions and theorems clear,
it should be noted that proposition~\ref{prp_no1} and theorem~\ref{gss_thm} are
also (implicitly) used in the proof of theorem~\ref{ptr_thm} whereas
theorem~\ref{gss_thm} is used in the proof of proposition~\ref{prp_no1}.

\subsubsection{Proof of Diophantus' 20th problem}

We start by assuming that $pq(q^2-p^2)$ is a square. Propositions~\ref{prp_no2}
and~\ref{prp_no4} allow us to claim that $p$, $q$ and $q^2-p^2$ are squares.
Let us have $q=m^2$, $p=n^2$ and $q^2-p^2=r^2$. Thus, we obtain:
\begin{equation}
r^2=q^2-p^2=m^4-n^4=(m^2+n^2)(m^2-n^2)
\label{dio_ini}
\end{equation}

We have:

\begin{itemize}
\item $m^2+n^2$ and $m^2-n^2$ are odd because $p$ and $q$ have distinct
parities;
\item $m$ and $n$ are relatively prime (proposition~\ref{prp_no3});
\item $m^2+n^2$ and $m^2-n^2$ are relatively prime (proposition~\ref{prp_no1}).
\end{itemize}

As $(m^2+n^2)(m^2-n^2)$ is a square, there exist (proposition~\ref{prp_no4})
two natural numbers $u$ and $v$ s.t.:
\begin{equation}
m^2+n^2=u^2\mx{ and }m^2-n^2=v^2
\label{uav_sys}
\end{equation}

But, $u^2=q+p$ and $v^2=q-p$. Then, $u$ and $v$ are odd and are relatively
prime. Moreover, $u^2-v^2=(u+v)(u-v)=2n^2$ and $u+v$, $u-v$ are even (divisible
by $2$). If $d$ is a common prime divisor of $u+v$ and $u-v$ then $d$ divides
$2u$ and $2v$ (by addition and subtraction). If $d>2$ then $d$ divides $u$ and
$v$ (theorem~\ref{gss_thm}): this leads to a contradiction because $u$ and $v$
are relatively prime. Thus, $\mathsf{gcd}(u+v,u-v)=2$.

However, the product of two even numbers is divisible by $4$. So, exactly one
of $u+v$ and $u-v$ is a multiple of $4$. Let us assume that $u-v$ is a multiple
of $4$: we have $u-v=4s$ and $u+v=2w$, with $s$, $w$ relatively prime and $w$
odd. Then we obtain:
$$(u+v)(u-v)=8sw=2n^2\mx{ and next: }n^2=4sw\Lra{}(\fc{n}{2})^2=sw$$

The numbers $s$ and $w$ are relatively prime and then $s$ and $w$ are squares
(proposition~\ref{prp_no4}). Thus, we have:
$$u-v=4a^2\mx{, }u+v=2b^2\mx{, }v=b^2-2a^2$$

Next:
$$n^2=4a^2b^2\mx{ and using }(\ref{uav_sys})\mx{: }m^2=n^2+v^2=b^4+4a^4$$

Writing $m^2=b^4+4a^4$ means that $(b^2,2a^2,m)$ is a Pythagorean triple (we
can remark that if we assume that $u+v$ is the multiple of $4$, we have the
same values for $m$ and $n$). We can express this triple as described by
theorem~\ref{ptr_thm} and observing that $b^2$ is odd (for $u$ and $v$ are
relatively prime):
$$(b^2,2a^2,m)=(k'(q'^2-p'^2),2k'p'q',k'(p'^2+q'^2))$$

It is necessary that $k'=1$ since $b^2$ and $2a^2$ are relatively prime (for
$u$ and $v$ are relatively prime) and we have:
$$b^2=q'^2-p'^2\mx{, }a^2=p'q'$$

Finally, for the same reason, $p'$ and $q'$ are also relatively prime. 
As $p'q'$ and $(p'+q')(q'-p')$ are squares, $p'$, $q'$, $p'+q'$ and $q'-p'$ are
also squares (proposition~\ref{prp_no4}). Setting $q'=m^2$ and $p'=n^2$, we are
back to the initial point: looking for $m^2$ and $n^2$ whose addition and
subtraction must be squares implies looking for $m'^2$ and $n'^2$ with the same
property. But we have $m'^2+n'^2<m^2+n^2$:
$$m'^2+n'^2=q'+p'=\fc{b^2}{(q'- p')}<b^2<b^2+2a^2<(b^2+2a^2)^2=m^2+n^2$$

We can restart the reasoning and we will always find strictly smaller
non-negative integers (not w.r.t. $m$ and $n$ but w.r.t. $m^2+n^2$) verifying
the same conditions. However, this leads to a contradiction because there does
not exist an infinity of smaller non-negative integers (bounded by $0$). This
reasoning was called {\em infinite descent} by Fermat. Thus, $pq(q^2-p^2)$
cannot be a square and Diophantus' 20th problem has no solution.

\subsection{Application: Fermat's last theorem for $n=4$}
\label{fer_thm}

From the proof of Diophantus' 20th problem, we can deduce quite directly the
proof of Fermat's last theorem for $n=4$, i.e. there do not exist three
non-zero natural numbers $x$, $y$ and $z$ s.t. $x^4+y^4=z^4$. Regarding this
proof, we essentially used~\cite{Fer4-proof}, but it can be also found
in~\cite{JI63}, \cite{HW79} and~\cite{DS93}.

As previously (for Diophantus' 20th problem), the idea is to deduce a
contradiction and the proof starts by assuming that there exist
$x,y,z\in{}\ns{}^*$ s.t.:
\begin{equation}
x^4+y^4=z^4
\label{fer_ini}
\end{equation}

We can assume that $y$ and $z$ are relatively prime. Otherwise if $d$ is the
gcd of $y$ and $z$, then $y=dy'$, $z=dz'$ and we have:
$$z^4-y^4=d^4(y'^4-z'^4)=x^4$$

Thus, $d$ divides $x$ and if $x=dx'$ then we have to prove:
$$x'^4+y'^4=z'^4$$

which is the initial equation~(\ref{fer_ini}) with $y'$ and $z'$ relatively
prime.

We can also assume that $y$ and $z$ have distinct parities. First, $y$ and $z$
cannot be both even because we have just assumed that they are relatively
prime. Next, let us show that $y$ and $z$ can be supposed not to be both odd.
Equation~(\ref{fer_ini}) can be written as follows:
$$(x^2)^2+(y^2)^2=(z^2)^2$$

Thus, $(x^2,y^2,z^2)$ is a Pythagorean triple. As a consequence of
theorem~\ref {ptr_thm}, one of the numbers $x^2$ and $y^2$ is even (of the form
$2mpq$). By symmetry of ${\cal T}$, we can assume that $y^2$ is even
(otherwise we have to permute the role of $x$ and $y$: we can show that $x$
and $z$ are also relatively prime and we apply the same reasoning which
follows). In this way, $x^2$ and $z^2$ are both odd (divided by an odd $m$);
otherwise, they are both even (divided by an even $m$) which contradicts the
assumption that $y$ and $z$ are relatively prime. So, we can assume that $y^2$
and $z^2$ have distinct parities, as well as $y$ and $z$.

Moreover, equation~(\ref{fer_ini}) is equivalent to:
$$z^4-y^4=(z^2+y^2)(z^2-y^2)=x^4=(x^2)^2$$

This new equation shows that the problem is now reduced to proving that the
expression $(z^2+y^2)(z^2-y^2)$ cannot be a square, with $y$, $z$ relatively
prime and having distinct parities. This has been already shown in
subsection~\ref{inf_des} when proving Diophantus' 20th problem with infinite
descent. More precisely, we are exactly in the conditions of
equation~(\ref{dio_ini}), where $m$, $n$ are relatively prime and have distinct
parities (since $p$ and $q$ have distinct parities).

\section{Formalization}

\subsection{Generalities}

As mentioned in the introduction, we used the \cq{} proof assistant (latest
version V8.0~\cite{CoqDocWeb8.0}) to carry out the entire formalization of
Diophantus' 20th problem. This choice was essentially motivated by some of the
recent improvements proposed by this version of \cq{}. Amongst other features,
we were attracted by the complete revision of the concrete syntax which appears
more homogeneous and which allows us to get a kind of overloading with a system
of scopes. In particular, for number theory, this is quite appropriate because
we have exactly the same notations (e.g. for $0$, $1$, $+$, $*$, etc) over
$\ns{}$, $\zs{}$, $\rl{}$ or $\rs{}$. Despite the fact that the proof style and
the level of automation provided by \cq{} is not as suitable as could be
expected for mathematical developments, this release does clearly represent a
step toward a more mathematician-friendly framework.

Regarding the formalization, it was also necessary to make some choices
essentially motivated by the developments provided by the standard library of
\cq{} as well as the level of automation offered by the system. For example,
as seen in section~\ref{mth_prf}, the theorem deals only with natural numbers
but we use many expressions with the opposite \verb+-+ (together with
appropriate side conditions ensuring that the corresponding expressions are
always natural numbers; see equation~(\ref{dio_nwf}), for example) and as
$\ns{}$ is only a semi-ring, the automation strategy over rings (tactic
\verb+Ring+) does not work as expected (it does not simplify expressions
involving the opposite). As a consequence, many algebraic simplifications must
be carried out manually using the appropriate combination of rewritings. This
tends to slow down the development significantly and we decided to use $\zs{}$
(with some additional non-negativity conditions) instead of $\ns{}$. In this
way, the theorem is formally the same and we get a full automation for
algebraic manipulations (the tactic \verb+Ring+ does work as expected). Another
point which had to be dealt with is that \cq{}'s standard library does not
provide a rational number theory (used in the proof of theorem~\ref{ptr_thm}).
Actually, there are several libraries of rationals (contributed by some \cq{}
users), but no standard tends to emerge and especially none of them is related
to the classical real number theory provided by the standard library. To work
around this problem, we considered the real number library and we used an {\em
ad hoc} rational predicate (considering that a rational number is a real number
expressed as a fraction of two integers), which was quite sufficient to deal
with our proof.

In the following, we present an outline of our formalization which has been
separated in three significant parts: the characterization of Pythagorean
triples, the application of infinite descent and the proof of Fermat's last
theorem for\linebreak{}$n=4$. The whole development is available as a \cq{}
contribution~\cite{CoqContribs8.0}. For information, this contribution involves
about 2000~lines of code and took the equivalent of two months of development.

\subsection{Pythagorean triples}

The proof in \cq{} of theorem~\ref{ptr_thm} follows exactly the steps described
in subsection~\ref{pyt_tri} (trying to characterize the non-negative rational
coordinates of the unit circle). We do not give all the intermediary lemmas
necessary to build the proof and here are the two main lemmas (step~6) which
allows us to conclude:

{\small
\begin{verbatim}
Lemma pytha_thm1 : forall a b c : Z,
  (is_pytha a b c) -> (pytha_set a b c).

Lemma pytha_thm2 : forall a b c : Z,
  (pytha_set a b c) -> (is_pytha a b c).
\end{verbatim}
}

where \verb+is_pytha+ is the Pythagorean triple predicate (corresponding to
${\cal S}$) and \verb+pytha_set+ is the set of Pythagorean triples
(corresponding to ${\cal T}$), which are defined as follows:

{\small
\begin{verbatim}
Definition pos_triple (a b c : Z) :=
  (a >= 0) /\ (b >= 0) /\ (c >= 0).

Definition is_pytha (a b c : Z) :=
  (pos_triple a b c) /\ a * a + b * b = c * c.

Definition cond_pqb (p q : Z) :=
  p >= 0 /\ q > 0 /\ p <= q /\ (rel_prime p q).

Definition distinct_parity (a b : Z) :=
   (Zeven a) /\ (Zodd b) \/ (Zodd a) /\ (Zeven b).

Definition cond_pq (p q : Z) := cond_pqb p q /\ (distinct_parity p q).

Definition pytha_set (a b c : Z) :=
  exists p : Z, exists q : Z, exists m : Z,
    (a = m * (q * q - p * p) /\ b = 2 * m * (p * q) \/
     a = 2 * m * (p * q) /\ b = m * (q * q - p * p)) /\
    c = m * (p * p + q * q) /\ m >= 0 /\ (cond_pq p q).
\end{verbatim}
}

where \verb+Z+ corresponds to $\zs{}$, \verb+Zeven+/\verb+Zodd+ are
respectively the even/odd predicates over \verb+Z+ (predefined in the \cq{}
library) and \verb+rel_prime+ is the relatively prime predicate over \verb+Z+
(also predefined).

\subsection{Infinite descent}

\subsubsection{Infinite descent and induction}

Historically, infinite descent~\cite{PSF22,AW84,JI63}, invented in the 17th
century by Fermat, is one of the first explicit uses of reasoning by
induction\footnote{Here, by induction, we mean {\em complete induction} (or {\em
mathematical induction}), in contrast to {\em incomplete induction}, which was
used in Fermat's time to establish conjectures and which simply consisted in
verifying the validity of a proposition over $\ns{}$ for the first values of
$\ns{}$.} (over natural numbers) in a mathematical proof (around the same time,
Blaise~Pascal used a similar principle to prove properties for numbers in {\em
his triangle}). Nevertheless, as claimed in~\cite{CPW04}, some tend to think
that this principle was, in fact, already used by the ancient Greeks (in
particular, by the Pythagorean mathematician Hippasos~of~Metapont in the proof
of the irrationality of the golden number $\fc{1}{2}(1+\sqrt{5})$) in the 5th
century BC, and thus, long before Fermat, who simply reinvented it. Formally,
Fermat's induction schema can be expressed in a general way as follows:
\begin{equation}
(\fa{}x.P(x)\rt{}\ex{}y.y\pe{}x\we{}P(y))\rt{}\fa{}x.\nt{}P(x)
\end{equation}

where the relation $\pe{}$ is supposed to be well-founded.

This schema is quite appropriate to establish universally false properties
(in particular, Diophantus' 20th problem) but even if it appears that Fermat
failed to adapt it to prove universally true properties\footnote{Actually, as
can be noticed in a work sent to Christiaan~Huygens {\em vîa} Pierre~de~Carcavi
(see~\cite{AW84,PSF22,JI63}), Fermat succeeded in using the descent to answer
positive questions, operating a kind of $\nt{}\nt{}$-translation over the
statement, more or less easily in some cases (for example, every prime number
of the form $4n+1$ is the sum of two squares) and quite painfully in some
others (such as, every number is a square or composed of two, three or four
squares). However, he never used a positive induction schema to do so.}, this
principle is, in fact, equivalent to Noetherian induction~\cite{TC-ind,CPW04},
which allows us to prove properties {\em positively} and which is the
following:
$$(\fa{}x.(\fa{}y.y\pe{}x\rt{}P(y))\rt{}P(x))\rt{}\fa{}x.P(x)$$

where the relation $\pe{}$ is supposed to be well-founded.

Thus, to apply one or the other of these schemas to our proof
(see subsection~\ref{inf_des}), we only have to prove that the relation
${\cal R}(x,y)(x',y')\eq{}x+y<x'+y'$ (over $\ns{}$) is well-founded. This is
trivially done using a compatibility lemma related to the relation $<$
(predefined in the \cq{} library), i.e. if there exists a function $f$ s.t.
${\cal R}(x,y)\rt{}f(x)<f(y)$ then ${\cal R}$ is well-founded. Here, in our
case, the function is simply $f(x,y)=x+y$.

\subsubsection{Development}

The formalization in \cq{} of Diophantus' 20th problem follows the steps
described in subsection~\ref{inf_des} and to conclude, we use the infinite
descent schema. As said previously, for the infinite descent principle, we
started proving the Noetherian induction lemma adapted to our proof (using
the well-foundedness induction schema provided by the library of \cq{}, as well
as the proof that the relation given previously is well-founded) and then we
deduced the infinite descent lemma. Here are some of the corresponding lemmas
(we proved the infinite descent schema for $\ns{}$ and we generalized it, with
non-negativity side conditions, to work over $\zs{}$):

{\small
\begin{verbatim}
Lemma noetherian : forall P : nat * nat -> Prop,
  (forall z : nat * nat, (forall y : nat * nat,
    (fst(y) + snd(y) < fst(z) + snd(z))%nat -> P y) -> P z) ->
  forall x : nat * nat, P x.

Lemma infinite_descent_nat : forall P : nat * nat -> Prop,
  (forall x : nat * nat, (P x -> exists y : nat * nat,
    (fst(y) + snd(y) < fst(x) + snd(x))%nat /\ P y)) ->
  forall x : nat * nat, ~(P x).

Lemma infinite_descent : forall P : Z -> Z -> Prop,
  (forall x1 x2 : Z, 0 <= x1 -> 0 <= x2 ->
    (P x1 x2 -> exists y1 : Z, exists y2 : Z, 0 <= y1 /\ 0 <= y2 /\
    y1 + y2 < x1 + x2 /\ P y1 y2)) ->
  forall x y: Z, 0 <= x -> 0 <= y -> ~(P x y).
\end{verbatim}
}

where the notation \verb+%nat+ is used to switch to the arithmetic scope of
\verb+nat+ (the default scope has been set for \verb+Z+), the symbol \verb+*+
is the Cartesian product and \verb+fst+/\verb+snd+ are respectively the
first/second components of a couple.

Next, here are four lemmas corresponding to the propositions stated in the
preliminaries of subsection~\ref{inf_des} (as said in this subsection, Gauss's
theorem has already been proved in \cq{} and is part of the standard library):

{\small
\begin{verbatim}
Lemma prop1 : forall m n : Z, rel_prime m n -> distinct_parity m n ->
  rel_prime (m + n) (m - n).

Lemma prop2 : forall m n : Z, rel_prime m n ->
  rel_prime (m * m) (n * n) /\ rel_prime m (m * m - n * n).

Lemma prop3 : forall m n : Z, rel_prime (m * m) (n * n) -> rel_prime m n.

Lemma prop4 : forall p q : Z, 0 <= p -> 0 <= q -> rel_prime p q ->
  is_sqr (p * q) -> is_sqr p /\ is_sqr q.
\end{verbatim}
}

where \verb+is_sqr+ is the square predicate defined as follows:

{\small
\begin{verbatim}
Definition is_sqr (n : Z) : Prop :=
  0 <= n -> exists i : Z, i * i = n /\ 0 <= i.
\end{verbatim}
}

Finally, here are the two main lemmas, a refined version of the problem (i.e.
looking for $p$, $q$ s.t. $pq(q^2-p^2)$ is a square) and the final problem:

{\small
\begin{verbatim}
Lemma diophantus20_refined : forall p q : Z,
  p > 0 -> q > 0 -> p <= q -> rel_prime p q -> distinct_parity p q ->
  ~is_sqr (p * (q * (q * q - p * p))).

Lemma diophantus20 :
  ~(exists x : Z, exists y : Z, exists z : Z, exists t : Z,
    0 < x /\ 0 < y /\ 0 < z /\ 0 < t /\ x * x + y * y = z * z /\
    x * y = 2 * (t * t)).
\end{verbatim}
}

\subsection{Fermat's last theorem for $n=4$}

The formalization in \cq{} of Fermat's last theorem for $n=4$ follows the proof
described in subsection~\ref{fer_thm}. As previously stated, the idea is to use
the refutation of equation~(\ref{dio_ini}), established by the descent in the
proof of Diophantus' 20th problem and expressed as follows:

{\small
\begin{verbatim}
Lemma diophantus20_equiv : forall y z : Z,
  y > 0 -> z > 0 -> y <= z -> rel_prime y z -> distinct_parity y z ->
  ~is_sqr ((z * z + y * y) * (z * z - y * y)).
\end{verbatim}
}

Here are the main lemma as well as a refined version making the application of
the previous lemma possible:

{\small
\begin{verbatim}
Lemma fermat4_weak:
   ~(exists x : Z, exists y : Z, exists z : Z,
      0 < x /\ 0 < y /\ 0 < z /\ rel_prime y z /\ distinct_parity y z /\
      x * x * x * x + y * y * y * y = z * z * z * z).

Lemma fermat4:
   ~(exists x : Z, exists y : Z, exists z : Z,
      0 < x /\ 0 < y /\ 0 < z /\
      x * x * x * x + y * y * y * y = z * z * z * z).
\end{verbatim}
}

\section{Conclusion}

\subsection{Related proofs and formalizations}

One of the most significant related proofs is certainly John~Harrison's work,
who did the same formalization in \hq{} (an old implementation of the
\hl{}~\cite{HOL} system). Actually, it is not exactly the same especially
regarding the proof of Pythagorean triples (theorem~\ref{ptr_thm}), which, as
seen in subsection~\ref{pyt_tri}, is based on the characterization of the
rational points of the unit circle. Moreover, the formalization described here
is fully constructive in contrast to Harrison's; we do not use the excluded
middle or any form of the axiom of choice (the real numbers we use are
classical but this could be avoided relying on a constructive formalization of
real numbers or more appropriately of rational numbers; unfortunately, none of
these formalizations are standard theories in \cq{}).

In \cq{}, some non trivial proofs regarding number theory have been also 
developed (as user contributions, see~\cite{CoqContribs8.0}). For example,
Olga~Caprotti and Martijn~Oostdijk formalized Pocklington's criterion for
checking primality for large natural numbers (their development includes also a
proof of Fermat's little theorem). Valérie~Ménissier-Morain also developed a
proof of Chinese lemma (related to the notion of congruence) and finally,
Laurent~Théry~\cite{LT03} formalized the correctness proof of Knuth's algorithm
which gives the first $n$ prime numbers.

In other theorem provers, the \mr{} system~\cite{MizPres} provides a large
library of formalizations (the \mr{} Mathematical Library). In particular, a
subset of this library is dedicated to Mathematics and is edited as the
collection entitled {\em Formalized Mathematics}~\cite{Miz-FM04}, which
contains many developments regarding number theory. In \hl{} (and variants),
Joe~Hurd~\cite{JH03} formalized the Miller-Rabin probabilistic primality test
and John~Harrison is developing the Agrawal-Kayal-Saxena primality test.
Finally, in \ie{}~\cite{Isa2003}, the project directed by
Jeremy~Avigad~\cite{Math-Isa} at Carnegie~Mellon~University aims at developing
Mathematics in \ie{}'s higher-order logic and is focusing, in particular, on
extending the number theory library of the \ie{} system.

\subsection{Extensions}

As far as the authors know, this work is one of the first formalizations
(together with Harrison's) of a proof based on the infinite descent principle
(other formalizations must certainly use Noetherian inductions but they are not
expressed in the infinite descent way). This opens up some possibilities of
re-using this method, which can be easily generalized to any well-founded
relation, for some other proofs which may be appropriate for this kind of
reasoning (essentially universally false properties). As examples, we have
another historical proof, which is the proof of Fermat's last theorem for
$n=3$~\cite{HW79} (which is, in fact, the basic case if we try to prove
Fermat's last theorem by induction). The proof (maybe by Fermat and later by
Euler and Gauss independently) also uses the principle of infinite descent but
is longer and far more technical than that for $n=4$. This should not be
considered as surprising: induction can be applied trivially in some proofs
whereas in some others, it turns out to be tricky to make it work and this is
also true for the infinite descent schema. Also, it would be possible to adapt
the method to formalize other proofs (equally historical) of the same theorem
for other specific values of $n$ ($n=5$, $n=7$, etc), which similarly use the
descent and which essentially come from attempts to prove the theorem in the
general case (in this situation, it may appear surprising that the breakthrough
came from a link with algebraic geometry and did not use any kind of
induction). But, more generally, as pointed out in~\cite{AW84}, infinite
descent is the method {\em par excellence} in number theory and in Diophantine
analysis. In this way, some other projects could be Fermat's
equation~\cite{DS93,HW79,AW84,JI63} (also wrongly called Pell's equation in
older writings; i.e. the equation $x^2-Ny^2=1$ has infinitely many solutions in
$\zs{}$ if $N>1$ and is not a square), where the method of descent could be
used to get a proof of existence (but not to compute solutions), or, more
ambitiously and also more {\em modern}, the proof of Mordell's
theorem~\cite{AW84} (the group of rational points of an elliptic curve is
always finitely generated), where the descent has been refined to be applied.


\begin{thebibliography}{10}

\bibitem{Math-Isa}
Jeremy Avigad, Kevin Donnelly, and Paul Raff.
\newblock Mathematics in {{\sf Isabelle}}, 2004.
\newblock \\{\sf http://www.andrew.cmu.edu/\~{}avigad/isabelle/}.

\bibitem{CoqContribs8.0}
The {\sf Coq}~User Community.
\newblock The {{\sf Coq}} {U}ser's {C}ontributions, January 2005.
\newblock \\{\sf http://coq.inria.fr/contribs-eng.html}.

\bibitem{TC-ind}
Thierry Coquand.
\newblock Inductive {D}efinitions and {T}ype {T}heory: an {I}ntroduction, 1999.
\newblock Preliminary draft for the TYPES Summer School. \\{\sf
  http:www.cs.chalmers.se/\~{}coquand/ind.ps}.

\bibitem{PSF22}
Pierre~Simon de~Fermat.
\newblock {\em \OE{}uvres complètes (4 vols.)}.
\newblock Éditées par Paul Tannery et Charles Henry, Gauthier-Villars, Paris,
  1894-1912.
\newblock Avec un supplément de C. de Waard, 1922.

\bibitem{KD94}
Keith~J. Devlin.
\newblock {\em Mathematics: the Science of Patterns}.
\newblock Scientific American Library. W.H. Freeman (New York), 1994.
\newblock ISBN 0-7167-5047-3.

\bibitem{Miz-FM04}
Formalized {M}athematics, 2004.
\newblock ISSN 1426-2630.\\{\sf http://mizar.uwb.edu.pl/JFM/index.html}.

\bibitem{HOL}
M.~J.~C. Gordon and T.~F. Melham.
\newblock {\em Introduction to {\sf HOL}: a Theorem Proving Environment for
  Higher Order Logic}.
\newblock Cambridge University Press, 1993.

\bibitem{HW79}
G.~H. Hardy and E.~M. Wright.
\newblock {\em An Introduction to the Theory of Numbers}.
\newblock Oxford University Press, London, UK, 5th edition, 1979.
\newblock ISBN 0198531702.

\bibitem{JH03}
Joe Hurd.
\newblock Verification of the {M}iller-{R}abin {P}robabilistic {P}rimality
  {T}est.
\newblock {\em Journal of Logic and Algebraic Programming}, 50(1--2):3--21,
  May--August 2003.
\newblock Special issue on Probabilistic Techniques for the Design and Analysis
  of Systems.

\bibitem{JI63}
Jean Itard.
\newblock {\em Arithmétique et théorie des nombres}, volume 1093 of {\em Que
  sais-je}.
\newblock P.U.F., 1963.

\bibitem{Dio-proof}
ChronoMath~(Serge Mehl).
\newblock Descente infinie selon {F}ermat, 2005.
\newblock \\{\sf http://serge.mehl.free.fr/anx/desc\_inf.html}.

\bibitem{Fer4-proof}
ChronoMath~(Serge Mehl).
\newblock Grand théorème de {F}ermat, cas n = 4, 2005.
\newblock \\{\sf http://serge.mehl.free.fr/anx/th\_ferm4.html}.

\bibitem{Dio-Ari}
Diophantus of~Alexandria.
\newblock Arithmetica, c. {\sc AD} 250.

\bibitem{Euc-Ele}
Euclid of~Alexandria.
\newblock Stoicheion, c. {\sc BC} 300.

\bibitem{Isa2003}
Larry Paulson and Tobias Nipkow.
\newblock The {{\sf Isabelle}} {H}ome {P}age, 2003.
\newblock \\{\sf http://www.cl.cam.ac.uk/Research/HVG/Isabelle/index.html}.

\bibitem{DS93}
Daniel Shanks.
\newblock {\em Solved and Unsolved Problems in Number Theory}.
\newblock Chelsea Publishing Co., Inc., New York, USA, 4th edition, 1993.
\newblock ISBN 0-8284-2297-X.

\bibitem{SS02}
Simon Singh.
\newblock {\em Fermat's Last Theorem}.
\newblock Fourth Estate, June 2002.
\newblock ISBN 1841157910.

\bibitem{CoqDocWeb8.0}
The Coq~Development Team.
\newblock {\em The Coq Proof Assistant Reference Manual Version~8.0}.
\newblock INRIA-Rocquencourt, January 2005.
\newblock \\{\sf http://coq.inria.fr/doc-eng.html}.

\bibitem{LT03}
Laurent Théry.
\newblock Proving {P}earl: {K}nuth's {A}lgorithm for {P}rime {N}umbers.
\newblock In {\em Proceeding of Theorem Proving in Higher Order Logics
  (TPHOLs), Rome (Italy)}, volume 2758 of {\em LNCS}, pages 304--318.
  Springer-Verlag, September 2003.

\bibitem{MizPres}
Andrzej Trybulec.
\newblock The {{\sf Mizar}}-{QC}/6000 {L}ogic {I}nformation {L}anguage.
\newblock In {\em ALLC Bulletin (Association for Literary and Linguistic
  Computing)}, volume~6, pages 136--140, 1978.

\bibitem{AW84}
André Weil.
\newblock {\em Number Theory: An Approach through History from Hammurapi to
  Legendre}.
\newblock Boston MA Basel Stuttgart: Birkhäuser, 1984.
\newblock ISBN 0-8176-3141-0.

\bibitem{CPW04}
Claus-Peter Wirth.
\newblock Descente infinie + {D}eduction.
\newblock In {\em Logic Journal of the IGPL}. Oxford University Press, 2004.

\end{thebibliography}
\end{document}